\documentclass{aanew}
\usepackage{psfig}

\def\bron{SAX J1808.4-3658}
\def\ecs{erg~cm$^{-2}$s$^{-1}$}
\def\lum{erg~s$^{-1}$}

\begin{document}

\title{The first outburst of \bron\ revisited}
\titlerunning{The first outburst of \bron\ revisited}
\authorrunning{J.J.M. in 't Zand, R. Cornelisse, E. Kuulkers et al.}

\author{J.J.M.~in~'t~Zand\inst{1,2}
 \and  R.~Cornelisse\inst{1,2}
 \and  E.~Kuulkers\inst{1,2}
 \and J.~Heise\inst{1}
 \and L.~Kuiper\inst{1}
 \and A.~Bazzano\inst{3}
 \and M.~Cocchi\inst{3} 
 \and J.M.~Muller\inst{1,4,}\thanks{Present address: University Hospital
      Nijmegen, Dept. of Radiology, Computer Aided Diagnosis, P.O. Box 9101,
      6500 HB Nijmegen, the Netherlands}
 \and L.~Natalucci\inst{3}
 \and M.J.S.~Smith\inst{1,5,}\thanks{Present address: XMM-Newton Science
      Operations Center, Vilspa Satellite Tracking Station, Apartado 50727,
      28080 Madrid, Spain}
 \and P.~Ubertini\inst{3} 
}

\offprints{J.J.M. in 't Zand, email {\tt jeanz@sron.nl}}

\institute{     Space Research Organization Netherlands, Sorbonnelaan 2,
                NL - 3584 CA Utrecht, the Netherlands 
	 \and
                Astronomical Institute, Utrecht University, P.O. Box 80000,
                NL - 3508 TA Utrecht, the Netherlands
         \and
                Istituto di Astrofisica Spaziale (CNR), Area Ricerca Roma Tor
                Vergata, Via del Fosso del Cavaliere, I - 00133 Roma, Italy
         \and
                BeppoSAX Science Data Center, Nuova Telespazio,
                Via Corcolle 19, I - 00131 Roma, Italy
         \and
                BeppoSAX Science Operation Center, Nuova Telespazio,
                Via Corcolle 19, I - 00131 Roma, Italy
	}

\date{Received, accepted }

\abstract{
Data of the 1996 outburst of the single-known accreting millisecond
pulsar \bron, taken with the Wide Field Cameras (WFCs) on {\em BeppoSAX}, are
revisited with more complete data coverage and more comprehensive analysis
techniques than in a previous report. 
An additional type-I X-ray burst was identified 
which occurred at a time when the persistent emission is below the detection
limit, roughly 30 days after outburst maximum. This burst is three
times longer than the first two bursts, and 50\% brighter. It is the brightest
burst within the $\sim$1700 type-I bursts detected so far with the WFCs.
A spectral analysis of the data reveals a distance to \bron\ of
$\sim$2.5~kpc. This is an update from a previously reported value of 4~kpc.
We present the evidence that we have for the
presence of oscillations at the pulsar frequency during part
of the newly found burst. Such an oscillation would lend
support to the idea that the frequency of millisecond burst
oscillations in other objects is very close to the neutron star rotation
frequency.
}

\maketitle 

\section{Introduction}
\label{intro}
\bron\ was discovered with the Wide Field Cameras (WFCs) onboard the
Italian-Dutch X-ray satellite {\em BeppoSAX}, in data taken during
September 1996 (In~'t~Zand et al. 1998). Two intense type-I X-ray bursts
were found which showed evidence of near-Eddington
luminosities. Thus, a distance was estimated of 4~kpc, and the
source was identified as a likely low-mass X-ray binary (LMXB). The
outburst lasted approximately 3 weeks above a detection threshold of
$10^{-10}$~\ecs\ (2-10 keV), as measured with the
All-Sky Monitor (ASM) on the Rossi X-ray Timing Explorer (RXTE).

The source was found with RXTE to be in outburst again in April-May 1998
(Marshall 1998).
First-time observations with the Proportional Counter Array
on RXTE showed a strong 401~Hz pulsar signal (Wijnands \& Van 
der Klis 1998). This represents the first and long-awaited
unambiguous detection of an accreting millisecond pulsar.
In the 2 years since this discovery, no other such pulsar has been found.
Chakrabarty \& Morgan (1998) were able to determine orbital parameters
from Doppler-shifted pulsations. The orbital period is 2.01~hr, and the mass
function of the secondary star 3.8$\times10^{-5}$~M$_\odot$. 
Radio emission was detected once, at the time when the X-ray was about to
disappear (Gaensler et al. 1999). The emission is attributed to ejection
of material from the binary. It is
believed that \bron\ is the proof for a link between LMXBs and radio
millisecond pulsars (e.g., Chakrabarty \& Morgan 1998). So far no
pulsations have been found at radio wavelengths. 

In January 2000, \bron\ was found to be in outburst for a third time
(Van der Klis et al. 2000).
Unfortunately, the Sun was too close to the source to allow detailed coverage
of the outburst. Nevertheless, the X-ray pulsar signal was again
found.

\begin{figure}[t]
\psfig{figure=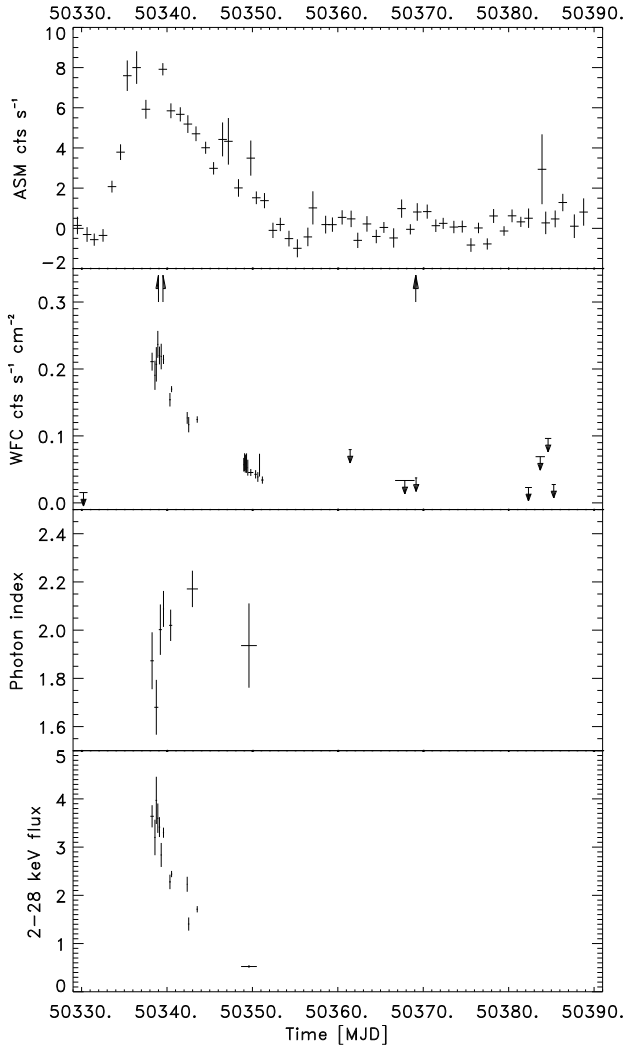,width=\columnwidth,clip=t}
\caption{Upper panel: ASM light curve of 2-12 keV daily averages during the
first outburst in 1996. In this time frame, there were
697 dwells over the source adding up to 63~ks. Second panel from above:
WFC 2-28 keV light curve
over the same period. The upward pointed arrows indicate the times of
three X-ray bursts detected with the WFC. The downward pointed arrows
indicate upper limits. The WFC exposure time in this time frame is 394~ks. For
a Crab-like spectrum, 1 WFC cts~s$^{-1}$cm$^{-2}$ is equivalent to 38
ASM cts~s$^{-1}$. Third panel: photon index resulting from WFC spectral 
fits to persistent
emission. Fourth panel: 2-28 keV flux resulting from WFC spectral fits,
in units of $10^{-9}$~\ecs.
\label{figasm}}
\end{figure}

\begin{figure}[t]
\psfig{figure=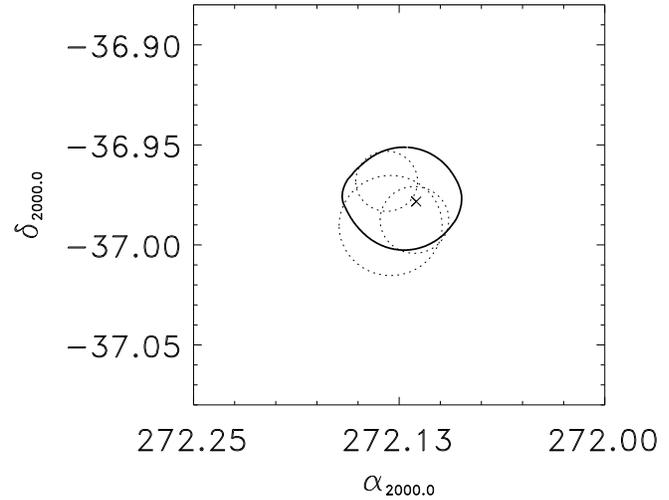,width=\columnwidth,clip=t}
\caption{Map of 99\%-confidence error regions of the 3 bursts (dotted
circles) with respect to the optical counterpart (cross) and the error
region of the persistent emission (thick loop).
\label{figwfcmap}}
\end{figure}

Because of its unique nature, \bron\ continues to be the subject of
scientific studies. In this circumstance we considered it appropriate
to revisit the WFC data of the first outburst. At the time of the first
report by In 't Zand et al. (1998), the millisecond pulsar was not yet
discovered and no high-resolution timing analysis was carried out on the
WFC data. Furthermore, the coverage of the processing of 1996 WFC data was
incomplete. Also, analysis techniques and calibration of the WFCs have
been updated to a more refined level. As a result of improvements, it is
now possible to obtain a more complete picture of
\bron\ during the first outburst. In this paper we present
a analysis of 'old' as well as 'new' data
and discuss the results. The main
new results are the detection of a third type-I X-ray burst, the marginal
detection of 401~Hz flux oscillations in that burst, and a
refinement of the distance to \bron. We review the WFC observations in
Sect.~\ref{observations}, discuss the data analysis in
Sects.~\ref{persistent},
\ref{analysis} and \ref{pds}, and discuss the implications in
Sect.~\ref{discussion}.

\section{Observations}
\label{observations}

The WFCs (Jager et al. 1997) are two identical coded aperture cameras, 
which were launched in April 1996 on board the {\em BeppoSAX} satellite
(Boella et al. 1997). Each camera has a field of view of 
$40^{\circ}\times40^{\circ}$
(full width to zero response), angular resolution of 5\arcmin\ and is
active between 2 and 28 keV. The on-axis detection threshold is of order a
few mCrab in $10^{5}$~s, and varies as a function of the total flux contained
in the field of view. In 1996 the Galactic center was targeted by the WFCs
for 22 days in the period August 21 to October 29. \bron\ is
9\fdg4 from the Galactic center, well within the field of view. The net 
exposure time for the source in 1996 is 550 ks.

Fig.~\ref{figasm} presents the time history of the flux as measured
with both WFC and ASM. The source was detected
with the ASM above a threshold of about 0.02 Crab units for 22 days starting
MJD~50333 (Sep. 5). The peak of 0.11 Crab units was reached within a 
few days. The WFC observations started while the source already was at the peak
emission for about 3 days and carried on intermittently till 1 month past
the moment when the source decayed below ASM detection levels.
The sensitivity varied considerably
due to varying off-axis angles over the different WFC observations. However,
typical sensitivities compare to that of the ASM measurements.

\begin{figure*}[t]
\psfig{figure=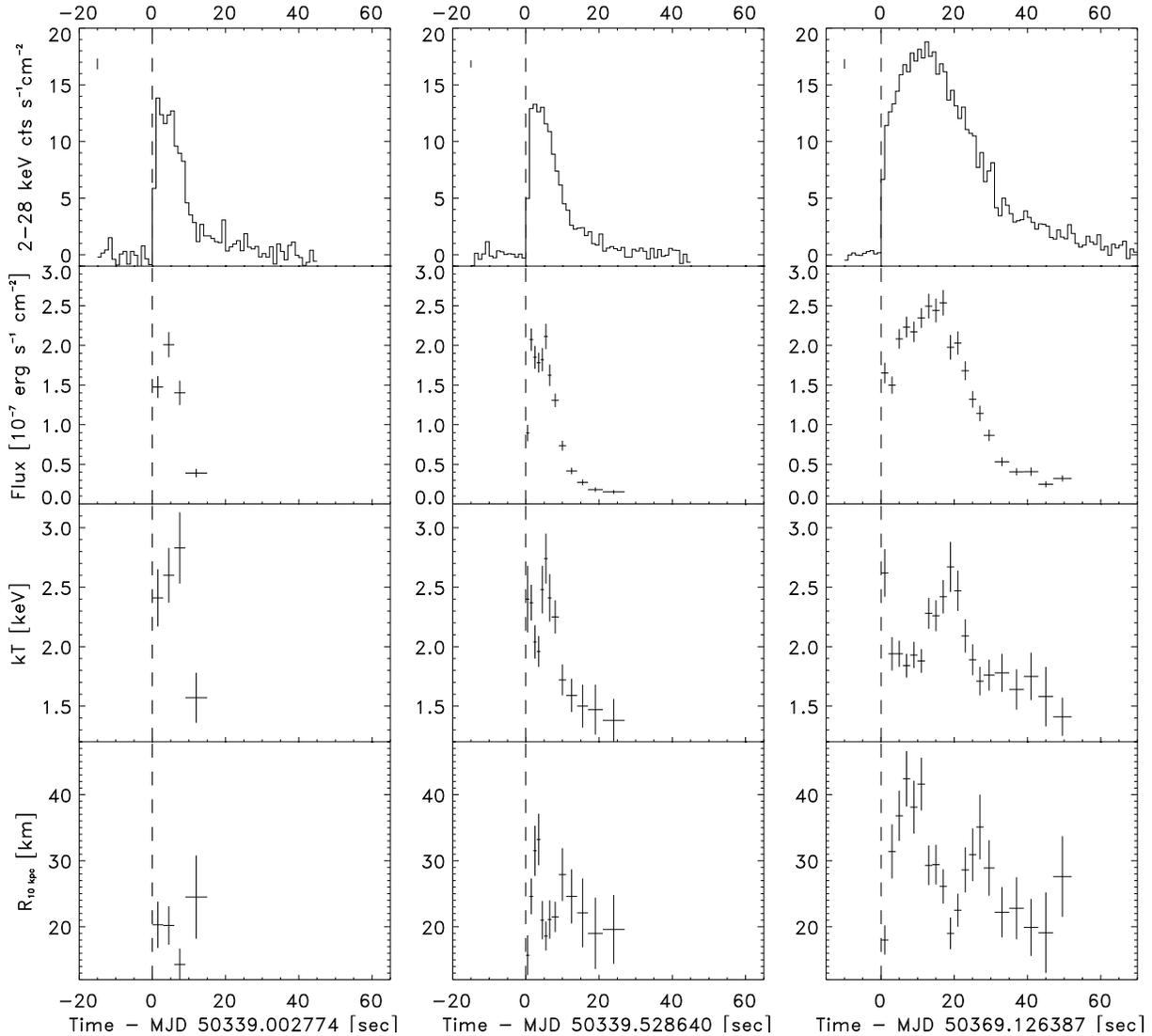,width=2\columnwidth,clip=t}
\caption{Time profiles of photon fluxes (upper
panel, with 1$\sigma$ error bars in the top left corner), bolometric flux
(second panel),
black-body color temperature (third panel) and black body radius (fourth panel)
of black body radiation spectral fits, for the three bursts detected
with WFC from \bron\ during the 1996 outburst. The vertical dashed 
lines are drawn for guiding purposes and mark the onsets of each burst
\label{figbursts}}
\end{figure*}

Fig.~\ref{figasm} also shows the times when X-ray bursts were detected. The
first two bursts have already been reported in In~'t~Zand et al. (1998), the
third one is for the first time reported here. The second burst follows
quite quickly the first one (0.5~d), while the third burst follows 30 days
later. Interestingly, the third burst occurred while no persistent emission
was measured. If the persistent spectrum would have been similar to
when the persistent flux was measured, the upper limit on the flux would
be $3\times10^{-10}$~\ecs\ (in 2-28 keV, for 10$^4$~s exposures).

Fig.~\ref{figwfcmap} shows the error regions of all bursts with respect to
that of the persistent emission and the optical counterpart (Giles
et al. 1998). There is no
doubt that these bursts are from \bron.

\section{Spectral analysis of the persistent emission}
\label{persistent}

The spectral data from the WFCs are read out in 31 channels between 2 and
28 keV. Three channels are not well calibrated and excluded in analyses. The
\bron\ detection data were resolved in 13 spectra. The data
could successfully be modeled by a simple power-law function with varying
photon index and a constant low-energy absorption, parameterized by
$N_{\rm H}$. For 7 different intervals, the resulting 
$\chi^2_{\rm r}=0.91$ (343 degrees of freedom). The photon index as a
function of time
is plotted in Fig.~\ref{figasm} (third panel). Although the data are not
consistent with a constant photon index ($\chi^2_{\rm r}=1.21$ for 349 dof),
no clear trend is apparent. $N_{\rm H}$ was undetermined below an upper limit
of $1\times10^{22}$~cm$^{-2}$. The outburst peak fluxes as measured with the
WFC are $4.0\times10^{-9}$~\ecs\ (2-28 keV), $2.1\times10^{-9}$~\ecs\ (2-10
keV) and $3.3\times10^{-9}$~\ecs\ (3-25 keV).

\section{Spectral analysis of bursts}
\label{analysis}

We carried out time-resolved spectroscopy of the three bursts
by modeling the 28-channel spectra with black body radiation.
The time resolution is a compromise between good statistics
to determine spectral parameters accurately enough, and time scales of
spectral changes. We used resolutions between 1 and 7~s.
The results are presented in Fig.~\ref{figbursts}. A number of 
observations can be made:
\begin{itemize}
\item all bursts exhibit a fast rise and exponential-like decay;
      they are consistent with black body radiation, the color
      temperature peaks in all three  cases at about 2.8~keV; and all 
      bursts show cooling in the tail. These characteristics clearly
      suggest that these are thermonuclear flashes on the surface of
      a neutron star (see also In~'t~Zand et al. 1998);
\item the second and third burst show variations in the extent of the
      emitting area and the corresponding color temperature,
      pointing to radius-expansion bursts due to luminosities near the
      Eddington limit. The asymptotic value for the extent of the emitting
      area, reached in the burst tails, is comparable for all three cases;
\item the third burst lasts about three times as long as the others. In terms
      of peak brightness (9.3 Crab units in 2-28 keV) it is 50\% brighter
      than the first two bursts. This makes the third burst the
      brightest type-I burst seen so far with the WFCs, out of a sample of
      1700 cases.
      In terms of bolometric fluxes the difference decreases to 20\%
      which is caused by a combination of a different color temperature 
      during the peak and the limited bandpass of the instrument. If
      correcting for gravitational redshift effects, the difference
      disappears (see below);
\item The bolometric fluences of the first two bursts are $2\times10^{-6}$
      and that of the third burst is $7\times10^{-6}$~erg~cm$^{-2}$.
      If no other bursts occurred between the first two bursts, 
      $\alpha$ (defined as the ratio of the fluence in the persistent
      emission between the bursts and that of the second burst) is
      $1.1\times10^2$. Given the ample WFC coverage, the wait time between
      bursts definitely is longer and $\alpha$ higher at later times
      when the third burst occurred.
\end{itemize}

\begin{figure}[t]
\psfig{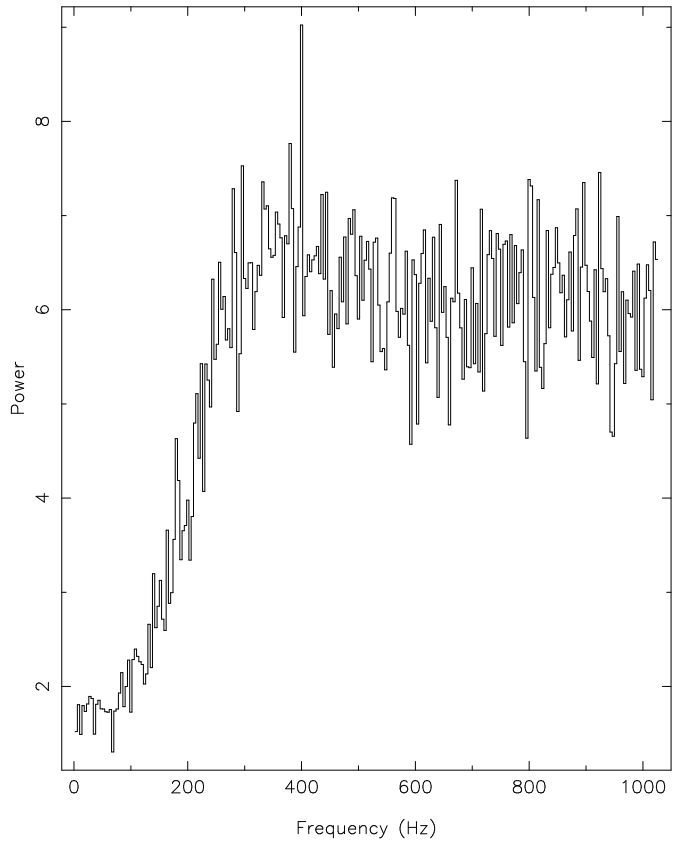}
\caption{PDS of the third burst. This is the average PDS of 80 data segments
of 0.25~s duration. The trends in
the background levels are due to the one-in-four event time labeling
for WFC data, as was verified with simulations.
The power is normalized following Leahy et al. (1983).
\label{figpds}}
\end{figure}

\section{Distance}

The distance to the source can be estimated from the assumption that during
the expansion and contraction phases the burst luminosity is close to the
Eddington luminosity (see e.g.\ Lewin et al.\ 1993). In principle this is
a simple method but in practice the uncertain value of the Eddington limit
makes it less straightforward. The Eddington limit depends on the mass of the
neutron star, on the electron scattering opacity (which changes with abundance
and temperature), on the degree of isotropy and on how close the
emission is to the neutron star (due to relativistic effects).
The latter effect introduces a dependency on the relation between the
color temperature to effective temperature. For a detailed discussion of these
dependencies, we refer to Kuulkers et al. (2001).

We determined the distance
to \bron\ by employing the five highest flux measurements of the 2nd burst and 
the seven highest of the 3rd burst.
Assuming standard burst parameters (isotropy, cosmic abundances,
1.4\,M$_{\odot}$ neutron star mass, equivalency of the color to effective
temperature, low-temperature opacity) we find distances of
$2.57\pm0.15$~kpc and $2.37\pm0.12$~kpc for the 2nd and 3rd burst,
respectively. If the
actual color to effective temperature correction is 1/1.7 instead of 1, the
distance would be about 10\% larger; if the hydrogen abundance is not
$X=0.73$ but $X=0.00$, the distance would be about 30\% larger; if the neutron
star mass is 2.0~M$_\odot$, the distance would be about 20\% larger. Therefore,
we conclude that the distance is 2.5~kpc but might just as well be 30\% larger.

The 2.5~kpc number is a revision of
the 4~kpc one reported by In~'t~Zand et al. (1998). This change is due to
1) application of the correction for gravitational redshift; 2)
rounding the final number off after two digits instead of one; and 3)
omitting the correction for detector dead time effects by In 't Zand et al.
(1998). All 3 effects contribute at comparable levels. We note that 
the dead time correction was not omitted in later WFC papers.

\section{Search for oscillations}
\label{pds}

In the telemetry data of the WFCs, each fourth event is labeled with
a time stamp of accuracy 2$^{-11}$~s and each 311th event with one
of accuracy 2$^{-12}$~s. This is
sufficient resolution to search for a coherent pulsation near 401~Hz.
If the total event rate is of the same magnitude or higher, the other
events (i.e., those without time stamps) may also contribute to the
sensitivity of such
a search because Poisson fluctations in the rate may occasionally
cause short enough wait times between two time stamps to enable 
accurate timing of the events in between.
We carried out
searches in two kinds of data: that of the bursts and that of the persistent
emission during the brightest
part of the transient outburst (on Sep. 12-13, 1996). The basis
for the searches are Fourier power density spectra (PDS) of light curves 
generated at 2$^{-11}$~s resolution from all photons that are
detected in those parts of the detector that are illuminated by \bron\
without being obstructed by the camera shielding or opaque parts of
the mask. For observations of \bron\ outside X-ray bursts, the total
event rate on the detector is between 620 and 770 s$^{-1}$.
The portion of the detector area illuminated by \bron\ varies
between 6 and 24\%. During the first 20~s of the third burst,
the average event rate is 1130 s$^{-1}$ and illuminated
detector portion 12\%.

It is technically not feasible to generate lightcurves from reconstructed
images of point sources at high time resolution. Therefore, we resort to taking
photons directly. This implies that the flux from other sources within 
20$^\circ$ from \bron\ along each image axis and within the field of view,
or from the background,
is not subtracted and the timing signatures of those sources will remain
present in the data. However, since we are searching for a very specific 
signature at around 401~Hz, this is not a major concern.

We barycentered the photon arrival times to the solar
system barycenter and included a correction for the binary orbit using the
ephemeris by Chakrabarty \& Morgan (1998) as determined from the 1998
outburst. This ephemeris is of sufficient accuracy to guarantee that the
orbital phase during the 1996 outburst is accurate to 10$^{-3}$. 

The burst data lengths were 10~s for the first 2 bursts and 30~s for the
third burst. While the first two bursts do not show any interesting signals,
a small signal was found near 401~Hz in one power spectrum of the third burst,
namely for
a frequency binning of 4~Hz (implied by a data segmenting in 0.25~s 
intervals), for the entire bandpass, and for averaging over the first 20~s of
data, see Fig.~\ref{figpds}. To assess the statistical significance of this
feature, we evaluated the probability of measuring such a power
or higher by chance through Monte Carlo (MC) calculations. We simulated
the circumstances
of the real observations, including the 1-in-4-events time stamping, the 
selection of the detector part and energy bandpass, the
burst light curve (see Fig.~\ref{figbursts}) and an instrumental
timing characteristic which causes a signal at 244~Hz and its
harmonics. No 401~Hz periodic signal was included. The continua
of the resulting power density spectra reproduce very well the observed one. 
Four out of 160,000 MC simulations resulted in powers higher than the one
observed. Thus, the chance probability for a single trial is
$2.5\times10^{-5}$. This is equivalent to a 4.0$\sigma$ detection.

The number of trials is determined by the number of times that
{\em meaningful} and {\em independent} searches for a burst oscillation
have been carried out.
This concerns three frequency binnings: 1, 2 and 4 Hz. Burst oscillations
often show a frequency drift of a few Hz (for a review, see, e.g., Strohmayer
1999). Therefore, we regard searches with a finer resolution than 1 Hz
as not meaningful. Searches are only meaningful for the second and third burst
because the statistical quality of the data of the first burst
prohibits detecting even the highest amplitudes. Searches were done
in two bandpasses, 2 to 10 and 10 to 28 keV (and the total band, but this
is a dependent trial). Finally, 
the sensitivity of a Fourier analysis that subdivides the data in a
number of time series of equal length, and
averages all resulting power spectra, is more sensitive when a number of
different phases are used when subdividing for a particular length. We
introduce 5 as the number of trials in this respect. In total, the
number of trials is 60. We note that in reality we carried out less trials
to find the signal, and that we did search for signals at higher frequency
resolution without positive results. The chance probability is 0.15\% for
60 trials. This is equivalent to a 3.0$\sigma$ detection.

\begin{figure}[t]
\psfig{figure=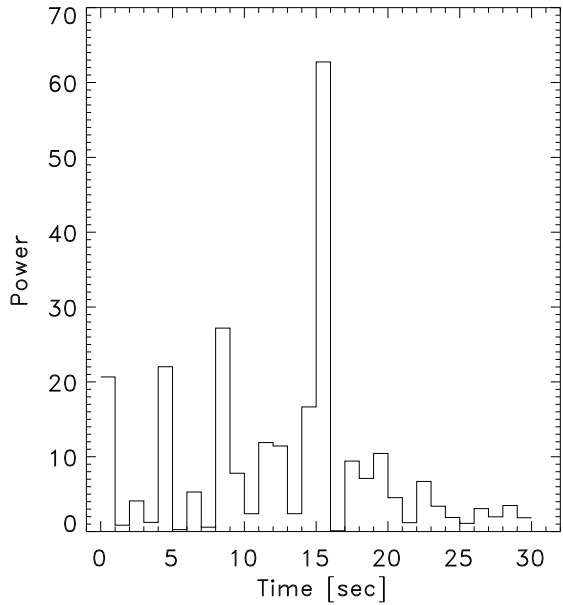,width=\columnwidth,clip=t}
\caption{Time history of power at 401~Hz, at 1~s time resolution 
and 1~Hz frequency resolution. Time is in s since burst onset. 
Same data as used in Fig.~\ref{figpds}.
The chance probability for detecting the peak power or higher is
10$^{-4}$ as was verified through MC calculations. The 10$^{-2}$
chance probability level is at a power of about 30.
\label{figpowerlc}}
\end{figure}

We tried to narrow down the time frame when the oscillation is present.
This is difficult with such a weak signal; nevertheless, when cutting
up the 20~s interval, the highest powers occur
toward the end of the interval. At 1~s and 1~Hz
resolutions, the peak occurs 15~s after the burst onset
(see Figs.~\ref{figpowerlc}). This is immediately after the radius
expansion phase, as is often seen in other burst oscillations.

The observed power is equivalent to an average oscillation amplitude
of 60$\pm6$\% of the source flux, as verified through simulations. If the
signal actually comes from a smaller time interval, its duration
is limited by a lower limit of 4~s. If the duration
would have been smaller, the implied amplitude would be larger than 100\%.

The data of the persistent emission has an elapse time of 139~ks and a net
exposure time of 32~ks. In
the resulting power density spectrum no coherent signal was found
within 20~Hz from 401~Hz. We also folded the data with the
400.9752106~Hz frequency as measured by Chakrabarty \& Morgan 1998 and
again found no significant signal. The upper limit on the amplitude
was determined
by simulating the circumstances of the observations and fitting folded
light curves with a sinusoid leaving free the amplitude and epoch. The
3$\sigma$ upper limit on the amplitude is about 20\% of the source flux.
This is a factor of 2 higher than the amplitude measured in the 1998
outburst (Wijnands \& Van der Klis 1998). Evidently, the pulsar signal is
too weak to be picked up in the persistent emission. 

\section{Discussion}
\label{discussion}

The evidence for $\sim$401~Hz burst oscillations in the third burst lends
support to the idea
that in other objects (with no pulsar signal) the frequency of burst
oscillations may be identified with the neutron star rotation frequency.
This is the first time that such direct evidence is presented for this idea.
However, we point out that the evidence is marginal.

The revision of the distance has as most important implication that
luminosities and intrinsic energy contents of outbursts are about 2.5 times
less than previously thought. The peak
2-28 keV luminosity of the first outburst is $3\times10^{36}$~\lum. If
the spectrum had the same power law to 100 keV, like 
in the second outburst (Gilfanov et al. 1998), the bolometric peak
luminosity may be of order $6\times10^{36}$~\lum, which is a few percent
of the Eddington luminosity for a canonical 1.4~M$_\odot$ neutron star.

The third burst is different from the other two bursts, in terms of
duration and fluence that are both about three times larger. The emitting
area, as measured with $R_{\rm 10~kpc}$ in the tail of the bursts, is the
same for all
3 bursts. Theory predicts (for reviews, see Lewin et al. 1993, Bildsten
1998, 2000) that the duration
and fluence of a type
I X-ray burst are mainly determined by the atomic composition of the
material being burned unstably during the burst, which is partly dependent on
the stable nuclear reaction rate taking place before the burst which, in turn,
depends on the accretion rate. The longer duration of the third
burst suggests mixed hydrogen/helium burning, which would
take place in a local accretion mass-flux regime of
$\dot{m}<900$~g~cm$^{-2}$s$^{-1}$ (Bildsten 2000). The persistent flux
during the third burst is measured to be below $5\times10^{35}$~\lum.
For a neutron
star radius of 5~km (at 2.5~kpc), this translates to an upper limit to
the mass flux rate of $9\times10^2$~g~cm$^{-2}$s$^{-1}$, which is
consistent. The first two bursts are at mass accretion rates one order
of magnitude above that limit, and are probably in the 
$\dot{m}>(2-5)\times10^3$~g~cm$^{-2}$s$^{-1}$ regime
of mixed hydrogen/helium bursts. This is confirmed by the value of $\alpha$.

Bursts were only seen in the first outburst. This can be explained
as a selection effect. If we assume that each outburst lasts 2 months,
the WFC observed the first outburst for 394~ks, the second outburst not
at all, and the third outburst for 34~ks. RXTE did not observe the first
outburst (except for 63~ks with the ASM), the second
outburst for 179~ks and the third outburst
for 146~ks. Ergo, the first outburst was covered the best.

\acknowledgement
We thank Michiel van der Klis and Fred Lamb for useful discussions,
and Jaap Schuurmans and Gerrit Wiersma for data archive and software
support. JZ and EK acknowledge financial support
from the Netherlands Organization for Scientific Research (NWO).
BeppoSAX is a joint Italian-Dutch program.

\end{document}